\documentstyle[preprint,aps,epsf]{revtex}
\newcommand {\bea}{\begin{eqnarray}}
\newcommand {\eea}{\end{eqnarray}}
\newcommand {\be}{\begin{equation}}
\newcommand {\ee}{\end{equation}}

\begin{document}

%\draft
\preprint{SUNY-NTG-00-14}

\title{Kaon Condensation in High Density Quark Matter}

\author{Thomas Sch\"afer}

\address{Department of Physics, SUNY Stony Brook,
Stony Brook, NY 11794\\ and\\
Riken-BNL Research Center, Brookhaven National 
Laboratory, Upton, NY 11973}

\maketitle

\begin{abstract}
  
 We point out that the problem of kaon condensation in 
dense hadronic matter can be addressed in perturbative 
QCD. Indeed, perturbative calculations suggest that negative
kaons are condensed in high density quark matter if the 
electroweak interaction is taken into account. This observation 
sheds new light on the proposal that the low density hyperon 
and high density quark matter phases of QCD are continuously 
connected.

\end{abstract}

\newpage

 The behavior of hadronic matter at very high baryon density has been 
a fascinating subject for quite some time. In the early 1970's Migdal, 
Sawyer, and Scalapino suggested that nuclear matter might exhibit pion 
condensation at densities near the saturation point of nuclear matter 
\cite{Migdal:1973,Sawyer:1972,Scalapino:1972}. In 1986 Kaplan and Nelson 
pointed out that kaons might condense at densities several times the 
saturation density \cite{Kaplan:1986,Brown:1987,Politzer:1991,Brown:1994}.
More recently, work on QCD at finite baryon density has mostly focussed 
on the behavior of quark matter at extremely high baryon density. This 
work goes back to the basic observation by Frautschi that asymptotic
freedom combined with the presence of a Fermi surface implies that cold 
quark matter is a color superconductor 
\cite{Frautschi:1978,Barrois:1979,Bailin:1984bm}. This idea was revived 
in \cite{Alford:1998zt,Rapp:1998zu} where it was emphasized that the 
corresponding gaps could be quite large, on the order of 100 MeV at 
densities 5-10 times larger than the nuclear saturation density. 

 The next important step was taken by Alford, Rajagopal,
and Wilczek who realized that in quark matter with
three flavors the dominant order parameter involves
the coupling of color and flavor degrees of freedom, 
``color-flavor-locking'' \cite{Alford:1999mk}
\be
\label{cfl}
\langle \psi_i^a C\gamma_5 \psi_j^b\rangle
 = \phi \left(\delta_i^a\delta_j^b-\delta_i^b\delta_j^a\right).
\ee
Here, $a,b$ are color indices and $i,j$ are flavor indices. 
This particular order parameter is distinguished by the
fact that it has the largest residual symmetry group
\cite{Alford:1999mk}, and therefore leads to the largest 
condensation energy \cite{Schafer:1999ef}. The 
color-flavor-locked condensate breaks both the original $SU(3)$ 
color symmetry and the $SU(3)_L\times SU(3)_R$ flavor symmetry,
but leaves a diagonal $SU(3)$ symmetry unbroken. This 
implies that chiral symmetry is spontaneously broken,
and that the spectrum contains almost massless 
pseudoscalar Goldstone bosons. Indeed, the excitation
spectrum of high density quark matter bears a remarkable
resemblance to the spectrum of low density hyperon 
matter. This has lead to the suggestion that the 
low and high density phases of three flavor QCD might 
be continuously connected \cite{Schafer:1999ef}.

 The systematic study of the low energy effective 
lagrangian of three flavor QCD at high density was
started by Casalbuoni and Gatto \cite{Casalbuoni:1999}. 
The effective lagrangian for the pseudoscalar Goldstone 
bosons takes the form \cite{Casalbuoni:1999,Son:1999}
\be
\label{leff}
{\cal L}_{eff} = \frac{f_\pi^2}{4} {\rm Tr}\left[
 \partial_0\Sigma\partial_0\Sigma^\dagger - v_\pi^2
 \partial_i\Sigma\partial_i\Sigma^\dagger \right]
 - c\left[\det(M){\rm Tr}(M^{-1}\Sigma) + h.c.\right].
\ee
Here, $\Sigma\in SU(3)$ is the Goldstone boson field, 
$v_\pi$ is the velocity of the Goldstone modes and 
$M={\rm diag}(m_u,m_d,m_s)$ is the quark mass matrix. 
The effective description is valid for energies and
momenta below the scale set by the gap, $\omega,q\ll
\Delta$. At very high density, $\mu\gg\Lambda_{QCD}$,
asymptotic freedom implies that the coupling is weak 
and the coefficients in the low energy lagrangian 
can be determined in perturbation theory. The first step 
is the calculation of the gap. The result for $N_f=3$ is
\cite{Son:1999uk,Schafer:1999jg,Pisarski:2000tv,Hong:2000fh,Brown:1999aq,Schafer:1999fe}
\be 
\label{gap}
 \Delta = 512\pi^4 2^{-1/3}(2/3)^{5/2}\mu g^{-5}
 \exp\left(-\frac{3\pi^2}{\sqrt{2}g}\right).
\ee  
Here, the factor $(2/3)^{5/2}$ reflects the larger 
amount of screening in three flavor QCD as compared 
to the two flavor case, and the factor $2^{-1/3}$ is
related to the structure of the color-flavor locked
state. 

 The low energy constants in (\ref{leff}) were 
determined by Son and Stephanov \cite{Son:1999},
see also \cite{Rho:1999,Hong:1999,Manuel:2000,Beane:2000}.
They find $v_\pi^2=1/3$ and
\bea
f_\pi^2 &=& \frac{21-8\log(2)}{18} \frac{\mu^2}{2\pi^2}, \\
 c &=& \frac{3\Delta^2}{2\pi^2}.
\eea
We can now determine the masses of the Goldstone 
bosons
\be 
\label{mgb}
 m^2_{\pi^\pm} = C(m_u+m_d)m_s, \hspace{1cm}
 m^2_{K_\pm}   = C m_d (m_u+m_s),
\ee
where $C=2c/f_\pi^2$. This result shows that the kaon is 
{\em lighter} than the pion. This can be understood from 
the fact that, at high density, it is more appropriate to 
think of the interpolating field $\Sigma$ as
\be
\label{sigma}
 \Sigma_{ij} \sim \epsilon_{ikl}\epsilon_{jmn}
 \epsilon^{abc}\epsilon^{dec}
 \bar\psi^a_{L,k}\bar\psi^b_{L,l}\psi^d_{R,m}\psi^e_{R,n}
\ee
rather than the more familiar $\Sigma_{ij}\sim
\bar\psi^a_{L,i}\psi^a_{R,j}$ \cite{Casalbuoni:1999}.
Using (\ref{sigma}) we observe that the negative pion 
field has the flavor structure $\bar{d}\bar{s}us$ and 
therefore has mass proportional to $(m_u+m_d)m_s$
\cite{Son:1999}. Putting in numerical values we
find that the kaon mass is very small, $m_{K^{-}} 
\simeq 5$ MeV at $\mu=500$ MeV and $m_{K^{-}}
\simeq 2.5$ MeV at $\mu=1000$ MeV. 

 We would like to remind the reader why this is so.
First of all, the Goldstone boson masses in the 
color-flavor-locked phase are proportional to the
quark masses squared rather than linear in the 
quark mass, as they are at zero density. This is 
due to an approximate $Z_2$ chiral symmetry in the 
color-flavor-locked phase \cite{Alford:1999mk}.
The diquark condensate is invariant under the 
transformations $\psi_{L,R}\to -\psi_{L,R}$, but
a linear Goldstone boson mass term is not. 
In addition to that, the Goldstone boson masses
are suppressed by a factor $\Delta/\mu$. This is
a consequence of the fact that the Goldstone modes
are collective excitations of particles and holes
near the Fermi surface, whereas the quark mass
term connects particles and anti-particles, far
away from the Fermi surface \cite{Rho:1999}. 

 The fact that the meson spectrum is inverted, and
that the kaon mass is exceptionally small opens 
the possibility that in dense quark matter electrons
decay into kaons, and a kaon condensate is formed. 
Consider a kaon condensate $\langle K^-\rangle =
v_Ke^{-i\mu_e t}$ where $\mu_e$ is the chemical 
potential for negative charge. The thermodynamic
potential ${\cal H}-\mu_e Q$ for this state is given 
by
\bea
\epsilon(\rho_q,x,y,\mu_e) &=&  \frac{3\pi^{2/3}}{4}
  \rho_q^{4/3} \left\{ x^{4/3} + y^{4/3}
 + (1-x-y)^{4/3} + \pi^{-4/3}\rho_q^{-2/3}m_s^2
  (1-x-y)^{2/3} \right\} \nonumber \\
 & & - \frac{1}{2}\left(\mu_e^2-m_K^2\right)v_K^2 +O(v_K^3)
 + \mu_e \rho_q \left(x-\frac{1}{3}\right) 
 - \frac{1}{12\pi^2}\mu_e^4
\label{eps}
\eea
Here, $\rho_q=3\rho_B$ is the quark density, and $x=\rho_u/
\rho_q$ and $y=\rho_d/\rho_q$ are the up and down quark 
fractions. For simplicity, we have dropped higher order terms 
in the strange quark mass and neglected the electron mass. 
These corrections are included in the results shown below. 
We have also assumed that neutrinos can leave the system. 
This assumption is appropriate in the case of neutron stars. 
In order to determine the ground state we have to make 
(\ref{eps}) stationary with respect to $x,y,\mu_e$ and $v_K$. 
Minimization with respect to $x$ and $y$ enforces $\beta$ 
equilibrium, while minimization with respect to $\mu_e$ ensures 
charge neutrality. Below the onset for kaon condensation 
we have $v_K=0$ and there is no kaon contribution to the
charge density. Neglecting $m_e$ and higher order corrections
in $m_s$ we find 
\be
 \mu_e\simeq \frac{m_s^2}{4p_F} 
  = \frac{m_s^2}{4\pi^{2/3}\rho_B^{1/3}}.
\ee
In the absence of kaon condensation, the electron chemical
potential will level off at the value of the electron mass
for very high baryon density. The onset of kaon condensation
is determined by the condition $\mu_e=m_K$. At this point it 
becomes favorable to convert electrons into negatively 
charged kaons. Once the amplitude of the kaon condensate 
starts to grow, nonlinear terms in the effective 
lagrangian have to be taken into account.

 Results for the electron chemical potential and the
kaon mass as a function of the light quark Fermi
momentum are shown in Fig.~1. In order to assess
some of the uncertainties we have varied the quark 
masses in the range $m_u=(3-5)$ MeV, $m_d=(6-8)$ MeV,
and $m_s=(120-150)$ MeV. We have used the one loop
result for the running coupling constant at two 
different scales $q=\mu$ and $q=\mu/2$. The scale
parameter was set to $\Lambda_{QCD}=238$ MeV,
corresponding to $\alpha_s(m_\tau)=0.35$ \cite{PDG}. 
An important constraint is provided by the condition
$m_s<\sqrt{2p_F\Delta}$ which ensures that flavor
symmetry breaking does not destroy color-flavor-locking
\cite{Alford:1999pa,Schafer:1999pb}. We have checked
that this condition is always satisfied for $p_F>
500$ MeV. Figure 1 shows that there is significant
uncertainty in the relative magnitude of the chemical
potential and the kaon mass. Nevertheless, the band 
of kaon mass predictions lies systematically below
the predicted chemical potentials. We therefore 
conclude that kaon condensation appears likely even 
for moderate Fermi momenta $p_F\simeq 500$ MeV. For
very large baryon density\footnote{This is not 
entirely correct. If $p_F>m_s^2/(4m_e)$ and kaons 
are not yet condensed then the system can no longer 
maintain $\beta$ equilibrium and $\mu_e$ goes to
zero.} $\mu_e\to m_e$ while $m_{K^-}\to 0$
and kaon condensation seems inevitable.  

 In the regime which is of physical interest the numerical 
values of $m_K$ and $\mu_e$ are very close, and it is important 
to address the uncertainties. 

 1. The most important uncertainty is related to the value of 
the gap. The leading terms in the perturbative expansion are
\be 
\label{pert}
\log\left(\frac{\Delta}{\mu}\right) =
-\frac{3\pi^2}{\sqrt{2}g} - 5\log(g)
 + \log\left( b_0^\prime 512\pi^4(2/N_f)^{5/2}\right) + \ldots .
\ee
While there is general agreement on the $O(g^{-1})$ and $O(\log(g))$ 
terms, the constant $b_0^\prime$ in the $O(g^0)$ term has not been 
completely determined yet. Brown et al. calculated the critical 
temperature up to order $O(g^0) $ and find $b_0^\prime =\exp(
-(\pi^2+4)(N_f-1)/16)\sim 0.17$ \cite{Brown:1999aq}, which corresponds 
to a substantial reduction of the gap. The origin of this correction
is a reduction of the strength of the quasiparticle pole in the 
dense medium. Manuel studied the effect of quasiparticle damping
and obtained a reduction of the gap by a factor $\sim 2$ at 
Fermi momenta $p_F\simeq 500$ MeV \cite{Manuel:2000nh}. This 
effect, however, appears to be a true higher order $O(g)$ 
correction. The modification of the gap due to color-flavor-locking
was considered in \cite{Schafer:1999fe,Shovkovy:1999mr,Evans:1999at}. 
This leads to a correction factor $2^{-1/3}$ which we have already 
included in (\ref{gap}). All these effects reduce the gap and increase
the likelihood of kaon condensation, at least as long as the 
magnitude of the gap exceeds the critical value for 
color-flavor-locking. 

 2. A related question is the problem of determining the 
scale $\Lambda$ at which the running coupling constant is
evaluated. This problem cannot really be addressed without
performing a higher order calculation. Beane et al. suggested 
to carry out a leading log resummation of the
gap equation \cite{Beane:2000yy}. This calculation results
in a substantial enhancement of the gap, corresponding to 
$b_0^\prime = \exp((33/16)(\pi^2/4-1))\sim 20$ in 
(\ref{pert}). This may be serious overestimate, however, 
because the authors use the perturbative beta function for 
momenta well below the screening scale $g\mu$. 

 3. At moderate densities QCD may generate a dynamical 
strange quark mass which is significantly bigger than 
the current quark mass $m_s\simeq 150$ MeV. This effect
helps kaon condensation because the electron chemical 
potential grows as $m_s^2$, whereas the kaon mass 
behaves as $m_s^{1/2}$. If the strange quark mass
becomes too large, then color is no longer locked 
to flavor and the kaon disappears. 

 4. Several authors have suggested that the Goldstone 
boson masses in the color-flavor-locked phase are of
the form $m_{GB}^2\sim m_q^2 (\Delta\bar\Delta/\mu^2)\log(\mu/
\Delta)$ \cite{Hong:1999,Beane:2000}, where $\bar\Delta$
is the ``gap'' for anti-particles. If $\bar\Delta=\Delta$
then the extra logarithm would lead to a modest increase 
of the kaon mass. Beane et al. also find a value of $f_\pi$ 
which is smaller, by a factor of 2, than the value quoted above. 
 
 5. Manuel and Tytgat considered the contribution of
a small instanton-generated quark condensate to the 
Goldstone boson masses \cite{Manuel:2000}. They conclude 
that this effect dominates over the perturbative 
result (\ref{mgb}) for chemical potentials of physical 
interest, $\mu<3$ GeV. We believe that this result is 
based on a mistake in the calculation of $\langle\bar
\psi\psi\rangle$ reported in \cite{Schafer:1999fe}. In that
work we calculated the effective quark mass generated
by instantons in the color-flavor-locked phase. Using
this result we extracted the quark condensate. In this
context we made a mistake similar to the one in the
first version of \cite{Son:1999}: Contrary to the 
result given in \cite{Schafer:1999fe} there is no
contribution to $\langle\bar{\psi}\psi\rangle$ which is 
proportional to the density of states on the Fermi
surface. The correct result is suppressed by an 
additional factor of $(\Delta/\mu)$. As a result,
terms linear in the quark mass may give an important 
contribution to the Goldstone boson masses in the case 
of moderate densities, $p_F\simeq 500$ MeV, but 
probably not for larger densities. 

 6. If the negative kaon mass becomes very small, the 
electromagnetic contribution to the mass may start to 
play a role. This effect was considered by Hong who finds
$\delta m^2_{GB^\pm}|_{em} \sim (e/g)^2 \Delta^2$, where the 
overall coefficient is of order 100 \cite{Hong:2000}. This 
result seems surprisingly large. A different estimate of 
the electromagnetic contribution to the charged pion and 
kaon masses can be obtained using a current algebra sum 
rule, which relates $\delta m^2_{GB^\pm}|_{em}$ to the 
difference between the vector and axial-vector spectral 
functions \cite{Das:1967}. The excitation spectrum in the 
vector channels was studied by Rho et al. \cite{Rho:2000}, 
who find that the vector and axial-vector are degenerate 
to leading order in perturbation theory. Using their results 
we conclude  that $\delta m^2_{GB^\pm}|_{em}<O(e^2\Delta^2
e^{-c/g})$. This estimate is oversimplified, because the 
derivation of the sum rule relies on Lorentz invariance. 
This question clearly deserves further study.

 Conclusions: We find that perturbative calculations 
suggest that negative kaons are condensed in high 
density baryonic matter. The mechanism for kaon condensation 
considered in this work is very different from the mechanism 
originally proposed by Kaplan and Nelson. In the original 
work, kaon condensation was driven by the attractive 
kaon-nucleon interaction implied by the kaon-nucleon
sigma term $\Sigma_{KN}$. In the color-flavor
locked phase all fermionic excitations have a large
gap and the interaction between Goldstone bosons
and quarks (baryons) is very weak. Kaon condensation
is possible because color-flavor locking inverts the
Goldstone boson spectrum and the kinematics of the 
Fermi surface suppresses Goldstone boson masses. The 
mechanism considered here shares an important feature
with the hadronic scenario of Brown et al. \cite{Brown:1994}
in that the presence of electrons is essential. 

 It is quite remarkable that the problem of kaon condensation
in dense hadronic matter can be addressed in perturbative 
QCD. This observation also sheds new light on the idea of 
quark-hadron continuity in QCD at finite baryon density. 
Even though low and high density QCD are qualitatively similar, 
there are important quantitative differences. Low density
QCD is characterized by large fermion masses and small 
Fermi surface gaps, whereas high density QCD has small
masses and large gaps. Also, while $f_\pi$ and 
$\langle(\bar\psi\psi)^2\rangle$ do not seem to
change much between $\mu=0$ and $\mu=500$ MeV, the 
quark condensate $\langle\bar\psi\psi\rangle$ is 
reduced by a large factor. We suggest that the 
low and high density phase are continuously connected,
and that the density above which it becomes more 
useful to describe matter in terms of quarks and
gluons rather than hadrons coincides with the 
point where the kaon and pion become degenerate. 

 We should note that a simple and robust mechanism 
for pion condensation in QCD with zero baryon
but finite isospin density, $\mu_u=-\mu_d$, was 
recently discussed by Son and Stephanov \cite{Son:2000xc}.

 Acknowledgements: This work was supported in part by US DOE grant 
DE-FG-88ER40388. I would like to acknowledge the hospitality of the 
European Center for Theoretical Studies in Nuclear Physics and
related Areas ECT* in Treto, Italy, where this work was completed.

\newpage\noindent

\begin{figure}
\begin{center}
\leavevmode
\vspace{1cm}
\epsfxsize=10cm
\epsffile{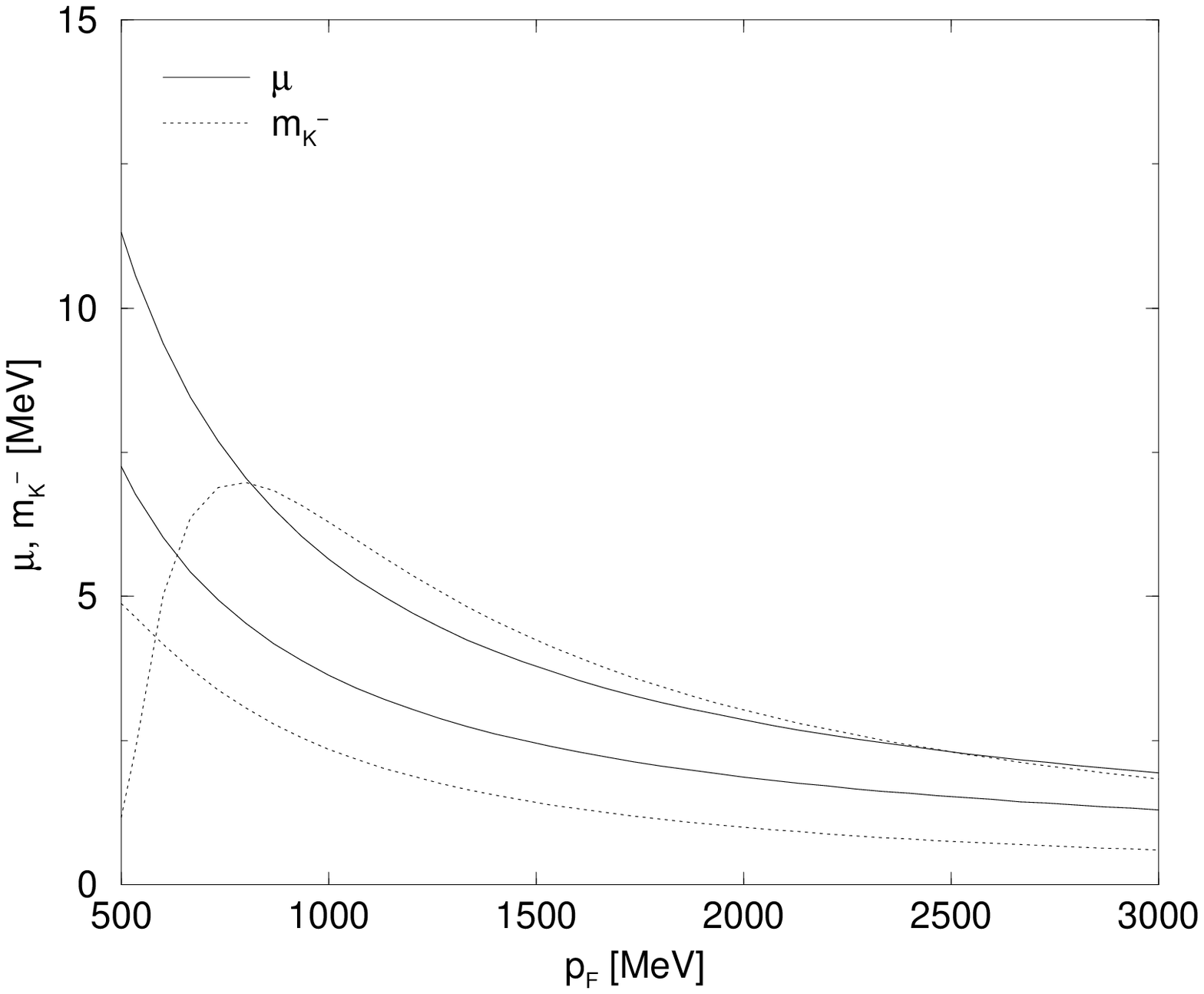}
\end{center}
\caption{Electron chemical potential (solid lines) and kaon mass
(dashed lines) in the color-flavor-locked quark phase. The two 
curves for both quantities represent a simple estimate of the 
uncertainties due to the value of the strange quark mass and
the scale setting procedure.}
\end{figure}

\end{document}